\documentclass[12pt]{article}
\usepackage{graphicx}
\usepackage{amssymb}
\usepackage{amsmath}

\newcommand{\bear}{\begin{array}}  
\newcommand {\eear}{\end{array}}
\newcommand{\bea}{\begin{eqnarray}}   
\newcommand{\eea}{\end{eqnarray}}
\newcommand{\beq}{\begin{eqnarray}}   
\newcommand{\eeq}{\end{eqnarray}}
\newcommand{\bef}{\begin{figure}}  \newcommand 
{\eef}{\end{figure}}
\newcommand{\bec}{\begin{center}}  \newcommand 
{\eec}{\end{center}}

\newcommand{\sla}[1]{\not \! \! #1}

\begin{document}

\begin{titlepage}

\begin{flushright}
IPMU10-0113\\
ICRR-Report-570-2010-3\\
\end{flushright}

\vskip 1.35cm

\begin{center}

{\large Gluon contribution to the dark matter direct detection}

\vskip 1.2cm

Junji Hisano$^{a,b}$, Koji Ishiwata$^c$, and Natsumi Nagata$^{a,d}$,

\vskip 0.4cm
{\it $^a$Department of Physics, Nagoya University, Nagoya 464-8602, Japan}\\
{\it $^b$Institute for the Physics and Mathematics of the Universe,
University of Tokyo, Kashiwa 277-8568, Japan}\\
{\it $^c$Institute for Cosmic Ray Research, University of Tokyo, 
Kashiwa 277-8582, Japan}\\
{\it $^d$Department of Physics, University of Tokyo, Tokyo 113-0033, Japan}\\

\date{\today}

\begin{abstract} 
  In this article we have calculated the spin-independent cross section of
  nucleon-dark matter scattering process at loop level, which is
  relevant to dark matter direct detection. Paying particular
  attention to the scattering of gluon with dark matter, which contributes
  as leading order in the perturbation, we have systematically
  evaluated loop diagrams with tracking the characteristic loop
  momentum which dominates in the loops. Here loop diagrams whose
  typical loop momentum scales are the masses of quarks and other
  heavier particles are separately presented. Then, we have properly
  taken into account each contribution to give the cross section.  We
  assume that the dark matter is pure bino or wino in the
  supersymmetric models. The application to other models is
  straightforward.

\end{abstract}



\end{center}
\end{titlepage}

\section{Introduction}

The existence of nonbaryonic dark matter has been established by
cosmological observations \cite{Dunkley:2008ie}.  Weakly-interacting
massive particles (WIMPs) are attractive candidates for dark matter,
and many models are proposed to predict WIMPs. The lightest neutralino
in the minimal supersymmetric standard model (MSSM) has been
extensively studied among them.

The direct detection experiments of WIMP dark matter are one of the
methods for probing the nature of the dark matter. In the experiments, one
searches for the signatures of WIMP-nucleon scattering. Many
experiments have searched for the dark matter signals, and their
sensitivities have been improved. The CDMS~II experiment
reported the final result for the five-tower WIMP search last
year~\cite{Ahmed:2009zw}. They observed two events in the dark matter
signal region for the detector; their data set the upper limit on the
spin-independent WIMP-nucleon elastic scattering cross section
(donated as $\sigma_{\rm SI}^N$) of $3.8\times 10^{-44}~$cm$^2$ for a
WIMP of mass 70~GeV.  This year the XENON100 experiment also started,
and has already reported the first result \cite{Aprile:2010um}; they
have given the upper limit as $\sigma_{\rm SI}^N<3.4\times
10^{-44}~$cm$^2$ for WIMPs of mass 55~GeV.  In addition, the XMASS
experiment will start soon \cite{Sekiya:2010bf}.  Prospects for
sensitivities of these new experiments are around $10^{-45}~$cm$^2$,
and they cover broad parameter space of the neutralino dark matter
scenario in the MSSM.

In order to study the nature of dark matter based on those latest
experiments, we need to evaluate the WIMP-nucleon elastic scattering
cross section precisely.  The WIMP-nucleon elastic scattering is
induced by effective interactions of the WIMP with quarks or gluon at
parton level. The effective interaction with the gluon is typically of
higher order of the QCD coupling constant than those with quarks in
many models, since the WIMP interacts with the gluon via loop diagrams
including quarks and/or other colored particles. However, the
effective scalar interaction with the gluon may make the leading
contribution in the spin-independent (SI) cross section. This comes
from the fact that the nucleon mass is dominated by gluon contribution.

The effective interaction of the WIMP with the gluon is generated by loop
diagrams.  The loop momentum which dominates the integration is
characterized by the mass scale of either heavy particles (such as WIMP mass)
or quarks. We call the former (the latter) the ``short-distance''
(``long-distance'') contribution. The long-distance contribution to the
effective scalar operator of the WIMP with gluon is approximately
evaluated from the effective scalar operators with quarks by using the
trace anomaly of QCD. On the other hand, the short-distance
contribution needs to be explicitly calculated. Drees and Nojiri
calculated the one-loop box diagrams for gluon-neutralino scattering
in the MSSM, and included the short-distance contribution to the SI
cross section \cite{Drees:1993bu}.  Although the SI cross
section is evaluated in various models, the short-distance
contribution is not included in many papers.

In this paper we show a method to include the short- and long-distance
contributions to the SI cross section systematically. We use the
Fock-Schwinger gauge for the gluon field. This is frequently used 
in the QCD sum rules in order to evaluate the operator product expansions
\cite{Novikov:1983gd}. It is found that this gauge makes the
calculation for the gluon contribution to the SI cross section much
more transparent. In addition, we separately calculate short- and
long-distance contributions, which makes it possible to add each
contribution properly to the cross section.

So as to study the details of gluon-neutralino scattering, we consider
two cases where the lightest neutralino is pure bino and wino in the
framework of the MSSM.  For the bino-like neutralino case, the
neutralino-quark scattering is induced by squark exchange at tree-level, 
and the neutralino-gluon scattering is generated at one-loop
level.  Calculating all the relevant diagrams at leading order,
consequently we have found that the long-distance contribution of
light quarks is partially regarded as the short-distance one in the
calculation as Ref.~\cite{Drees:1993bu}. This correction changes the
SI cross section slightly in the MSSM.  On the other hand, for the
wino-like neutralino case, when superparticles except for the
$SU(2)_L$ gauginos and the heavier Higgs bosons are decoupled in the
MSSM, the neutralino-nucleon scattering process is dominated by weak
gauge boson loop diagrams \cite{Hisano:2004pv}. We evaluated the
effective interactions of the neutralino with quarks at one-loop level
and those with gluon at two-loop level. Here we use the Fock-Schwinger
gauge for the gluon field, again; the result has been published in
Ref.~\cite{Hisano:2010fy}. In this paper, we show detail of the
evaluation.

Our method to evaluate the effective interaction of the WIMP with 
the gluon is
applicable to other models of dark matter, such as universal
extra-dimension scenario. Such analysis will be given elsewhere
\cite{HisanoIshiwataNagata}.

This article is organized as follows. In
Sec.~\ref{sec:effectivecoupling} we summarize the effective couplings
of the WIMP with quarks and the gluon. We assume that the WIMP is Majorana
fermion. Here we explain how to evaluate the effective interaction of
the WIMP with the gluon. In Sec.~\ref{sec:Bino}, assuming that the neutralino
is bino-like, we evaluate the effective couplings of the neutralino
with quarks and the gluon by using the Fock-Schwinger gauge for the gluon
field. Then we consider the case that the neutralino is wino-like
where the neutralino-nucleon scattering is generated at loop-diagrams
due to the weak interaction in Sec.~\ref{sec:Wino}.  Finally
Sec.~\ref{sec:conc} is devoted to the conclusion. In the Appendix, we show the
propagators of scalar and fermion in a gluon background
(Fock-Schwinger gauge) and other useful formulas for evaluation of
effective interaction of the WIMP with the gluon.

\section{Effective interaction  for WIMP-nucleon scattering}
\label{sec:effectivecoupling}

First, we summarize the effective interactions of the WIMP with light
quarks ($q=u,d,s$) and gluon, which are relevant to the WIMP-nucleon
spin-independent scattering. They are given as follows,
\begin{eqnarray}
{\cal L}^{\rm eff}
&=&\sum_{q=u,d,s}{\cal L}^{\rm{eff}}_q +{\cal L}^{\rm{eff}}_g \ ,
\end{eqnarray}
where 
\beq 
{\cal L}^{\rm{eff}}_q
&=& 
f_q m_q\ \bar{\tilde{\chi}}\tilde{\chi}\ \bar{q}q 
+ \frac{g^{(1)}_q}{M} \ \bar{\tilde{\chi}} i \partial^{\mu}\gamma^{\nu} 
\tilde{\chi} \ {\cal O}_{\mu\nu}^q
+ \frac{g^{(2)}_q}{M^2}\
\bar{\tilde{\chi}}(i \partial^{\mu})(i \partial^{\nu})
\tilde{\chi} \ {\cal O}_{\mu\nu}^q \
,
\label{eff_lagq}
\\
{\cal L}^{\rm eff}_{ g}&=&
f_G\ \bar{\tilde{\chi}}\tilde{\chi} G_{\mu\nu}^aG^{a\mu\nu}
+\frac{g^{(1)}_G}{M}\
 \bar{\tilde{\chi}} i \partial^{\mu}\gamma^{\nu}
\tilde{\chi} \ {\cal O}_{\mu\nu}^g
+
\frac{g^{(2)}_G}{M^2}\
\bar{\tilde{\chi}}(i\partial^{\mu}) (i\partial^{\nu})\tilde{\chi} 
\
{\cal O}_{\mu\nu}^g \ .
\label{eff_lagg}
\eeq
$\tilde{\chi}$ is the WIMP which we assume to be Majorana fermion as it is 
mentioned in the Introduction. $M$ and $m_q$ are masses for the WIMP and quarks,
respectively.  The second and third terms in ${\cal L}^{\rm eff}_q$
and ${\cal L}^{\rm eff}_g$ depend on the twist-2 operators (traceless
parts of the energy-momentum tensor) for quarks and gluon,
respectively,
\beq {\cal O}_{\mu\nu}^q&\equiv&\frac12 \bar{q} i
\left(D_{\mu}\gamma_{\nu} +
  D_{\nu}\gamma_{\mu} -\frac{1}{2}g_{\mu\nu}\sla{D}
\right) q \ ,
\nonumber\\
{\cal O}_{\mu\nu}^g&\equiv&\left(G_{\mu}^{a\rho}G_{\rho\nu}^{a}+
  \frac{1}{4}g_{\mu\nu} G^a_{\alpha\beta}G^{a\alpha\beta}\right) \ .
\label{twist2}
\eeq
Here the covariant derivative is defined as $D_\mu\equiv\partial_\mu+i
g_sA^a_\mu T_a$ ($g_s$, $T_a$ and $A^a_\mu$ are the $SU(3)_C$ coupling
constant and generator, and the gluon field, respectively), and
$G^a_{\mu \nu}$ is the field strength tensor of gluon.

The SI cross section of the WIMP with target nuclei $T$ is expressed compactly
in terms of the SI coupling of the neutralino with nucleon
$f_N~(N=p,n)$ \cite{Jungman:1995df};
\begin{eqnarray}
  \sigma^T_{\rm SI}&=&
  \frac{4}{\pi}\left(\frac{M m_T}{M +m_T}\right)^2
\left|n_p f_p+n_nf_n\right|^2\ , 
\label{sigma}
\end{eqnarray}
where $m_T$ is the mass of target nucleus, and $n_p$ and $n_n$ are
proton and neutron numbers in the target nucleus, respectively.  The
SI coupling of the neutralino with nucleon is given by the
coefficients and matrix elements of the effective operators in ${\cal
  L}^{\rm eff}_{q}$ and ${\cal L}^{\rm eff}_{g}$ as\footnote{
We have included contributions from effective interactions 
with twist-2 operators for charm and bottom quarks,
$g_c^{(i)}$ and $g_b^{(i)}$ $(i=1,2)$,  in Eq.~(\ref{f}).
See later discussion. 
}
\begin{eqnarray}
f_N/m_N&=&\sum_{q=u,d,s}
f_q f_{Tq}
+\sum_{q=u,d,s,c,b}
\frac{3}{4} \left(q(2)+\bar{q}(2)\right)\left(g_q^{(1)}+g_q^{(2)}\right)
\nonumber\\
&-&\frac{8\pi}{9\alpha_s}f_{TG} f_G 
+\frac{3}{4} G(2)\left(g^{(1)}_G
+g^{(2)}_G\right) \ . 
\label{f}
\end{eqnarray}
The matrix elements of the effective operators are expressed by
using nucleon mass as
\begin{eqnarray}
 \langle N \vert m_q \bar{q} q \vert N\rangle/m_N  &\equiv& f_{Tq}\ ,
\nonumber
\\
 1-\sum_{u,d,s}f_{Tq} &\equiv& f_{TG}  \ ,
\nonumber\\
\langle N(p)\vert 
{\cal O}_{\mu\nu}^q
\vert N(p) \rangle 
&=&\frac{1}{m_N}
(p_{\mu}p_{\nu}-\frac{1}{4}m^2_N g_{\mu\nu})\
(q(2)+\bar{q}(2)) \ ,
\nonumber\\
\langle N(p) \vert 
{\cal O}_{\mu\nu}^g
\vert N(p) \rangle
& =& \frac{1}{m_N}
(p_{\mu}p_{\nu}-\frac{1}{4}m^2_N g_{\mu\nu})\ 
G(2) \ .
\end{eqnarray}
In the matrix elements of twist-2 operators, $q(2)$, $\bar{q}(2)$ and
$G(2)$ are the second moments of the parton distribution functions
(PDFs) of quark, antiquark and gluon, respectively, 
\begin{eqnarray}
q(2)+ \bar{q}(2) &=&\int^{1}_{0} dx ~x~ [q(x)+\bar{q}(x)] \ ,
\cr
G(2) &=&\int^{1}_{0} dx ~x ~g(x) \ .
\end{eqnarray}

For the numerical calculation, we describe the input parameters for
the hadronic matrix elements that we use in this article. The mass
fractions of light quarks, $f_{Tq}$, are expressed as
\cite{Corsetti:2000yq}
\begin{eqnarray}
  f_{Tu}&=&\frac{m_u}{m_u+m_d}(1\pm\xi)\frac{\sigma_{\pi N}}{m_N}\ ,\nonumber\\
  f_{Td}&=&\frac{m_d}{m_u+m_d}(1\mp\xi)\frac{\sigma_{\pi N}}{m_N}\  ,\nonumber\\
  f_{Ts}&=&\frac{m_s}{m_u+m_d}y\frac{\sigma_{\pi N}}{m_N}\ 
\end{eqnarray}
for proton/neutron. Here, 
\begin{eqnarray}
\sigma_{\pi N}&=&\frac{m_u+m_d}2 \langle p|\bar{u}u+\bar{d}d|p\rangle  \ 
,\nonumber\\
y &=&\frac{2 \langle p|\bar{s}s|p\rangle}{\langle p|\bar{u}u
+\bar{d}d|p\rangle }\ , \nonumber\\
\xi&=&\frac{\langle p|\bar{u}u-\bar{d}d|p\rangle}
{\langle p|\bar{u}u+\bar{d}d|p\rangle}\ .
\end{eqnarray}
The recent lattice simulation predicts that $\sigma_{\pi N}= (53\pm
2({\rm stat})^{+21}_{-7} ({\rm syst}))~{\rm MeV}$ and $y =
0.030\pm0.016 ({\rm stat})^{+0.006}_{-0.008} ({\rm
  extrap})^{+0.001}_{-0.002} (m_s) $ \cite{Ohki:2008ff}. Combined with
$\xi =0.132\pm 0.035$ \cite{Corsetti:2000yq,Cheng:1988im}, we get
$f_{Tu}$= 0.023, $f_{Td}$= 0.032, and $f_{Ts}$=0.020 for the proton, and
$f_{Tu}$=0.017, $f_{Td}$= 0.041, and $f_{Ts}$= 0.020 for the neutron.  In
our numerical evaluation, we use the center value for each parameters.

The second moments of the PDFs of quark, antiquark and gluon are
scale-dependent, and are mixed with each other once the QCD radiative
corrections are included. Thus, the coefficients of the terms with
twist-2 operators become scale-independent after they are multiplied
by the second moments of the PDFs \cite{Drees:1993bu}. We use the
second moments for the PDFs at the scale of $Z$ boson mass, and
include bottom and charm quark contributions, as given in
Eq.~(\ref{f}).  The second moments are evaluated from the CTEQ parton
distribution \cite{Pumplin:2002vw} as
\begin{eqnarray}
G(2)=0.48,&& \nonumber\\
u(2)=0.22,&&\bar{u}(2)=0.034\ ,  \nonumber\\
d(2)=0.11,&&\bar{d}(2)=0.036\ ,  \nonumber\\
s(2)=0.026,&&\bar{s}(2)=0.026\ ,  \nonumber\\
c(2)=0.019,&&\bar{c}(2)=0.019\ ,  \nonumber\\
b(2)=0.012,&&\bar{b}(2)=0.012\ , 
\end{eqnarray}
for the proton. Those for the neutron are given by an exchange of up and down
quarks.  

For the calculation of $f_N$ at leading order, we give some remarks.
The effective interactions of the WIMP with the gluon are of higher order
of $\alpha_s (= g_s^2/4 \pi)$ than those of the light quarks in many
models including the MSSM. However, it is found in Eq.~(\ref{f}) that
the effective scalar coupling of the WIMP with the gluon $f_G$ gives the
leading contribution to the cross section even if it is suppressed by
one-loop factor compared with those of light quarks. On the other
hand, the contributions with the twist-2 operators of the gluon are of the
next-leading as $g^{(1)}_G$ and $g^{(2)}_G$ are suppressed by
$\alpha_s$.  Then, we ignore them in this paper.

Finally we explain the evaluation of $f_G$. In general, the effective
coupling of $\tilde{\chi}$-$g$ scattering is induced by loop diagrams
in which virtual quarks and other heavy particles run (like
squarks in the MSSM).  Consequently, the integral momentum around not
only the quark mass scale but also the heavy particle one
contributes.  (As we described in the Introduction, we call them
  as ``long-'' and ``short-distance'' contributions, respectively.)
Thus, in order to take into account both contributions accurately for
general cases, we must calculate loop diagrams explicitly. So as to
calculate diagrams including the quark systematically, we separate the two
contributions as
\begin{eqnarray}
f_G|_q  = f_G|^{\rm LD}_q + f_G|^{\rm SD}_q.
\end{eqnarray}
In the long-distance contribution, there is another approximated way to
evaluate the heavy quark $(Q=c,b,t)$ contribution, which is valid in the
limit of large mass of heavy particles.  After integrating out heavier
particles than heavy quarks, we write down the effective Lagrangian for
heavy quarks: ${\cal L}^{\rm eff}_Q = f_Q m_Q
\bar{\tilde{\chi}}\tilde{\chi} \bar{Q}Q+ \cdots$\footnote
{ We used the equation of motion for the quark to reduce the effective
  Lagrangian.  Though the heavy quarks are in loops in the evaluation
  of the effective interaction of the WIMP with the gluon, it is justified to
  use it.  See Ref.~\cite{Skiba:2010xn} and references in it. We also
  explicitly checked that the effective interaction
  $\bar{\tilde{\chi}}\tilde{\chi}\ \bar{Q} i\sla{D}Q$ leads to the
  same result for $f_G$ as $m_Q\bar{\tilde{\chi}}\tilde{\chi}\
  \bar{Q}Q$.  }.
We calculate one-loop triangle diagrams in which heavy quarks rotate to
emit two gluons, and get\footnote
{This result is also obtained in another way.  It is pointed out in
  Ref.~\cite{Drees:1993bu} that the effective interactions which come
  from heavy quark loops are approximately calculated in the effective
  interaction of the WIMP with heavy quarks, using the trace anomaly of
  energy-momentum tensor in QCD \cite{Shifman:1978zn}.  When the virtual
  heavy quark is the only particle in the loop diagram, the integrated-out
  heavy quark converts to the gluon operator as $ m_Q \bar{Q}Q \rightarrow
  -\frac{\alpha_s}{12 \pi}G_{\mu\nu}^a G^{a\mu\nu}$, which gives the
  same result as Eq.~(\ref{matching}).}
\begin{eqnarray}
f_G|_Q = -\frac{\alpha_s}{12 \pi}f_Q\ .
\label{matching}
\end{eqnarray}
(See Appendix for detail.) Thus, we expect that 
\begin{eqnarray}
f_G|^{\rm LD}_Q \simeq -\frac{\alpha_s}{12 \pi} f_Q\ ,
\label{eq:checkmatching}
\end{eqnarray}
in the limit of the large mass of heavy particles.  (This can be utilized
to check explicit loop calculation. See the following sections.)
In the explicit loop calculation, we observe that the momentum with the
scale of quark mass dominates the integral to give a factor $1/m_{Q}$
(which is canceled in $f_Q m_Q \bar{\tilde{\chi}}\chi \bar{Q}Q$).

For light quark contributions in long-distance loops, on the contrary,
we must not include them in the same manner as heavy quark ones when we
evaluate $f_G$.  As we saw in the above calculation, the loop momentum is
dominated by quark mass; thus, the propagators for the soft quark whose
momentum and mass are smaller than the QCD scale are determined
entirely by confinement dynamics.  The corresponding effect should be
included in the matrix element $\langle N| \bar{q}q |N \rangle$
\cite{Novikov:1983gd}.  That is why the light quark loops must not be
counted in the evaluation of $f_G$.  

Being aware of this fact, we sum up for quarks properly and get
\begin{eqnarray}
f_G = \sum_{q={\rm all}}f_G^{\rm SD}|_q + \sum_{Q=c,b,t}f_G^{\rm LD}|_Q\ .
\label{eq:fG}
\end{eqnarray}
In addition, it is pointed out \cite{Djouadi:2000ck} that the
long-distance contributions to $f_G$ suffer from large QCD correction.
The corrections to Eq.~(\ref{matching}) are given as 
\begin{eqnarray}
f_G|_Q =
-\frac{\alpha_s}{12\pi} c_Q f_Q\ ,
\label{matching2}
\end{eqnarray}
where $c_Q=1+11\alpha_s(m_Q)/4\pi$. We take $c_c=1.32$ and $c_b=1.19$
for $\alpha_s(m_Z)=0.118$ ($m_Z$ is $Z$ boson mass), and $c_t=1$ in
our numerical calculation.  Then, rewriting Eq.~(\ref{eq:fG}) with the
QCD correction, we finally get
\begin{eqnarray}
f_G = \sum_{q={\rm all}}f_G^{\rm SD}|_q + \sum_{Q=c,b,t}c_Q f_G^{\rm LD}|_Q\ .
\label{eq:fG2}
\end{eqnarray}

In the case where the WIMP interacts with the Higgs boson, the
short-distance contribution to the effective interaction of the WIMP with
the gluon is not generated at the leading order.  Then, we can easily
evaluate the SI cross section without explicit calculation of loop
diagrams by using Eq.~(\ref{matching2}).  However, it is not the case
for the bino-like neutralino dark matter.  In the next section, we
consider the case where the lightest neutralino is almost pure bino in
the MSSM.  The bino interacts with quarks and squarks. We will see that the
squark-quark loop diagrams give both short- and long-distance
contributions.

\section{Bino-like neutralino dark matter}
\label{sec:Bino}

The lightest neutralino interacts with quarks and gluon in diagrams
with Higgs boson exchange or squark exchange. When the
Higgsino-gaugino mixing is not negligible, the Higgs boson exchange
tends to dominate in the SI $\tilde{\chi}$-$N$ scattering. On the
other hand, when the mixing is negligible and squark masses are
comparable to the neutralino mass, the squark exchange dominates over
the Higgs boson exchange.  In this section, we take a limit $\mu \gg M$
($\mu$ is the Higgsino mass parameter in superpotential) and consider the
case where the lightest neutralino is almost pure bino. The extension to
the general cases is straightforward.

Before going to the numerical calculation, we show the couplings of
the effective Lagrangian which are given by the explicit calculation
of tree- and loop-level scattering amplitudes.  The following calculation
is applicable to the models in which the WIMP interacts with quarks
and new colored scalars.  (For the time being, we do not distinguish a
light or heavy quark for ``$q$'' unless it is expressly described, and
provide the analytic results of the effective couplings.)

\begin{figure}[t]
 \begin{center}
   \includegraphics[width=0.3\linewidth]{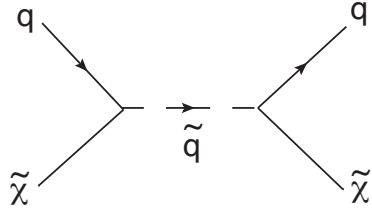} 
   \caption{Tree-level diagrams of squark ($\tilde{q}$) exchange to
     generate interaction of WIMP and light quark. }
\label{fig:tree}
 \end{center}
\end{figure}

First, we simply parametrize the neutralino interaction with quark and
squark $\tilde{q}$ as
\begin{eqnarray}
{\cal L}&=& \bar{q} (a_q+b_q \gamma_5) \tilde{\chi} \tilde{q}+{\rm h.c.}
\end{eqnarray}
Integrating out squarks at tree level (the corresponding diagram is
shown in Fig.~\ref{fig:tree}), the effective interaction with light
quarks in Eq.~(\ref{eff_lagq}) is generated as
\begin{eqnarray}
f_q&=&\frac{M}{(m_{\tilde{q}}^2-M^2)^2}\frac{a_q^2+b_q^2}8
-\frac1{(m_{\tilde{q}}^2-M^2)m_q}\frac{a_q^2-b_q^2}4  \ ,\nonumber\\
g_q^{(1)}&=&\frac{{M}}{(m_{\tilde{q}}^2-M^2)^2}\frac{a_q^2+b_q^2}2 \ ,\nonumber\\
g_q^{(2)}&=&0 \ . 
\label{treesq}
\end{eqnarray}
Here we take the zero quark mass limit.

\begin{figure}[t]
 \begin{center}
   \includegraphics[width=0.5\linewidth]{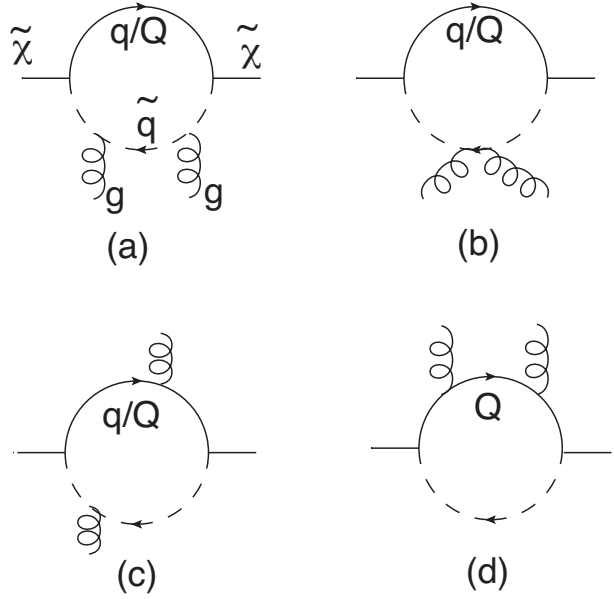} 
   \caption{One-loop diagrams to generate interaction of the WIMP and
     gluon. Here, $q$ and $Q$ are light and heavy quarks,
     respectively, and $\tilde{q}$ is for colored scalars, such as the squark
     in the MSSM.}
\label{fermion_dm}
 \end{center}
\end{figure}

At loop level, the interactions of the WIMP with gluon are generated.
The leading contribution is from one-loop diagrams including each
heavy or light quark in Fig.~\ref{fermion_dm}.  For the calculation of
these diagrams, we adopt the Fock-Schwinger gauge for the gluon field:
\begin{eqnarray}
x^{\mu}A_{\mu}^a(x) = 0.
\end{eqnarray}
By using this gauge, we can express the gauge field by the field strength
tensor as
\begin{eqnarray}
A_\mu^a(x)
&=&
\frac{1}{2} x^\rho G^a_{\rho\mu}(0)
+ \cdots \ ,
\end{eqnarray}
so that we can extract out relevant effective interaction of the gluon
easily. On the other hand, since the translation invariance is lost in
this gauge, we have to distinguish two propagators for quarks in a 
gluon background,
\begin{eqnarray}
i S(p) &\equiv& \int d^4 x e^{i p x} \langle 
T\{q(x)\bar{q}(0)\}\rangle)\ , \nonumber\\
i \tilde{S}(p) &\equiv& \int 
d^4 x e^{-i p x} \langle T\{q(0)\bar{q}(x)\}\rangle
\ ,
\end{eqnarray}
 and those for squarks,
\begin{eqnarray}
i \Delta(p) &\equiv& 
\int d^4 x e^{i p x} \langle T\{\tilde{q}(x)\tilde{q}^\dagger(0)\}\rangle
\ , \nonumber\\
i \tilde{\Delta}(p) &\equiv& \int 
d^4 x e^{-i p x} 
\langle T\{\tilde{q}(0)\tilde{q}^\dagger(x)\}\rangle
\ .
\end{eqnarray}
In the Appendix, propagators of the colored fermion and boson in
the Fock-Schwinger gauge are collected for convenience.  

The two-point function of the neutralino in the gluon background
(donated as $\Gamma_{\tilde{\chi}}$) is calculated by using the above
propagators; then the effective scalar coupling of neutralino and
gluon is derived from the two-point function as
\begin{eqnarray}
f_G|_q = \frac{ \Gamma_{\tilde{\chi}}(p)}{2} |_{GG}\ ,
\end{eqnarray}
where 
\begin{eqnarray}
i\Gamma(p) &=&
\int \frac{d^4q}{(2\pi)^4} 
\left[(a_q-b_q\gamma_5) S(p+q) (a_q+b_q \gamma_5)\right] \tilde{\Delta}(q)
\nonumber\\
&&+
 \int \frac{d^4q}{(2\pi)^4} 
\left[(a_q+b_q\gamma_5) \tilde{S}(p-q) (a_q-b_q \gamma_5)\right] {\Delta}(q)
\ .
\end{eqnarray}
Here, the factor $1/2$ comes from the neutralino being Majorana and 
``$|_{GG}$''
means to extract the coefficient for terms proportional to
$G_{\mu\nu}^aG^{a\mu\nu}$.  For simplicity, we write $\Gamma(p)$ to
which the single flavor of quark (and squark) contributes.  As we described in 
the previous section, we decompose it into short- and long-distance
contributions and write as 
\begin{eqnarray}
f_G^{\rm SD}|_q&=&
\frac{\alpha_s}{4\pi} 
\left(
\frac{a_q^2+b_q^2}4
M f^{s}_+
+\frac{a_q^2-b_q^2}4
m_q
  f^{s}_-
\right) \ , 
\nonumber\\
f_G^{\rm LD}|_q&=&
\frac{\alpha_s}{4\pi} 
\left(
\frac{a_q^2+b_q^2}4
M f^{l}_+
+\frac{a_q^2-b_q^2}4
m_q
  f^{l}_-
\right)  \ .
\label{fsdfld}
\end{eqnarray}
%
The short-distance contribution comes from diagram (b) in
Fig.~\ref{fermion_dm}, while the long-distance one is from diagram
(d). The diagrams (a) and (c) vanish in our gauge. (See Appendix.) The
mass functions are given as
\begin{eqnarray} 
f^{s}_+ &=& m^2_{\tilde{q}}~(B_0^{(1,4)}+B_1^{(1,4)})\ , \nonumber\\
f^{s}_- &=& m^2_{\tilde{q}}~B_0^{(1,4)}\ , \nonumber\\
f^{l}_+ &=& m^2_{q} ~(B_0^{(4,1)}+B_1^{(4,1)})\ , \nonumber\\
f^{l}_- &=& B_0^{(3,1)}+m^2_{q} ~B_0^{(4,1)}\ , 
\label{bfunc}
\end{eqnarray}
where
\begin{eqnarray}
  \int \frac{d^4q}{i \pi^2} 
  \frac{1}{((p+q)^2-m_q^2)^n(q^2-m_{\tilde{q}}^2)^m} 
  &\equiv&B_0^{(n,m)} \ ,\nonumber\\
  \int \frac{d^4q}{i \pi^2} 
  \frac{q_\mu}{((p+q)^2-m_q^2)^n(q^2-m_{\tilde{q}}^2)^m} 
  &\equiv&p_\mu B_1^{(n,m)} \ .
\label{bfuncdif}
\end{eqnarray}
In the above expression, it is apparent that the
typical momentum scale of integral is dominated by the squark mass in
the short-distance contribution while it is dominated by quark mass in
the long-distance one. With straightforward calculation of the integral,
we have finally obtained 
\begin{eqnarray}
  f^{s}_+&=&
  -\frac{(\Delta -6 m_{\tilde{q}}^2 m_q^2) (m_{\tilde{q}}^2+m_q^2-M^2)}
  {6 \Delta ^2 m_{\tilde{q}}^2}
  -\frac{2 m_{\tilde{q}}^2 m_q^4}{\Delta ^2}L \ ,
  \nonumber\\
  f^{s}_-&=&
  - \frac{3\Delta  m_{\tilde{q}}^2-(\Delta -6 m_{\tilde{q}}^2 m_q^2)
    (m_{\tilde{q}}^2-m_q^2+M^2)}{6 \Delta ^2 m_{\tilde{q}}^2}
  +\frac{ m_{\tilde{q}}^2 m_q^2 (m_{\tilde{q}}^2-m_q^2-M^2)}{\Delta ^2}L \ ,
  \nonumber\\ 
  f^{l}_+&=&
  -\frac{\Delta +12 m_{\tilde{q}}^2 m_q^2}{6 \Delta ^2}+
  \frac{m_{\tilde{q}}^2 m_q^2 (m_{\tilde{q}}^2+m_q^2-M^2)}{\Delta ^2} L \ ,
  \nonumber\\
  f^{l}_-&=&
  \frac{3\Delta  m_q^2+2(\Delta+3 m_{\tilde{q}}^2 m_q^2) 
    (m_{\tilde{q}}^2-m_q^2-M^2)}{6 \Delta ^2 m_q^2}-\frac{m_{\tilde{q}}^2 
    (\Delta +m_q^2 (m_{\tilde{q}}^2-m_q^2+M^2))}{\Delta ^2}L \ .
  \nonumber\\
\end{eqnarray}
Here, 
\begin{eqnarray}
  \Delta&=&
  M^4-2 M^2(m_q^2+m_{\tilde{q}}^2)+ (m_{\tilde{q}}^2-m_q^2)^2 \ ,
  \nonumber\\
  L&=& 
\left\{
\begin{array}{l}
  \frac1{\sqrt{|\Delta|}} \log\frac{m_{\tilde{q}}^2+m_q^2-M^2
    +\sqrt{|\Delta|}}{m_{\tilde{q}}^2+m_q^2-M^2-\sqrt{|\Delta|}} ~~~(\Delta>0) \ ,
\\
  \frac2{\sqrt{|\Delta|}} \tan^{-1} \frac{\sqrt{|\Delta|}}
  {m_{\tilde{q}}^2+m_q^2-M^2}
  ~~~(\Delta<0) \ .
\end{array}\right.
\end{eqnarray}
When we take zero quark mass limit, the mass functions are approximated
as
\begin{eqnarray}
f^{s}_+&\simeq& -\frac1{6 m_{\tilde{q}}^2(m_{\tilde{q}}^2-M^2)}  \ ,
\nonumber\\
f^{s}_-&\simeq& -\frac{2 m_{\tilde{q}}^2-M^2}
{6 m_{\tilde{q}}^2(m_{\tilde{q}}^2-M^2)^2} \ ,
\nonumber\\
f^{l}_+&\simeq& -\frac1{6 (m_{\tilde{q}}^2-M^2)^2} \ ,
\nonumber\\
f^{l}_-&\simeq& \frac1{3 m_q^2(m_{\tilde{q}}^2-M^2)} \ .
\end{eqnarray} 
For the heavy quark loop in the long-distance contribution, it is checked
that Eq.~(\ref{eq:checkmatching}) is satisfied from
Eqs.~(\ref{matching}) and (\ref{treesq}), just replacing $q$ as $Q$,
when $m_{\tilde{q}}-M\gg m_Q$, as expected. The mass function
$f^{l}_+$ in the long-distance contribution is not singular in a limit
of zero quark mass, while $f^{l}_-$ is. This is because  $f^{l}_+$ is
proportional to $m_q^2$ before the loop momentum integral and the loop
momentum integral leads to $1/m_q^2$.

With all results, we obtain the effective coupling of the neutralino with
gluon.  As we described in the previous section, the contribution of light
quarks should not be included in $f_G$ since the loop momentum has
infrared cutoff, $\sim$ GeV, implicitly.  In Ref.~\cite{Drees:1993bu},
however, the long-distance contribution of light quarks, {\it i.e.}
$f_G^{\rm LD}|_q~(q=u,d,s)$, is added in $f_G$.  This point can be
seen by comparing our result with theirs in a simple expression
under a certain limit. In the case of $m_{\tilde{q}}\gg M$, for
instance, our result gives
\begin{eqnarray}
  f_G \simeq \frac{\alpha_s}{4 \pi} \sum_{q={\rm all}}
  \left(
    -\frac{a_q^2+b_q^2}{24}\frac{M}{m_{\tilde{q}}^4} 
  \right)
  + \frac{\alpha_s}{4 \pi} \sum_{Q=c,b,t} 
  \left(
    -\frac{a^2_Q+b^2_Q}{24} \frac{M}{m^4_{\tilde{q}}}
    +\frac{a^2_Q-b^2_Q}{12} \frac{1}{m_Q m^2_{\tilde{q}}}
  \right) \ .
\end{eqnarray}
Focusing on the term which is proportional to $a^2_q+b^2_q$ for the
comparison, the result given in Ref.~\cite{Drees:1993bu} has the
form in which the term, $-\alpha_s/4 \pi \sum_{q=u,d,s}
(a_q^2+b^2_b)M/24m^4_{\tilde{q}}$, is added to the above
expression.
We found that the long-distance contribution of light quarks,
proportional to $a_q^2+b_q^2$, is regarded as the short-distance one
in Ref.~\cite{Drees:1993bu} and added in $f_G$.

\begin{figure}[t]
 \begin{center}
   \includegraphics[width=0.6\linewidth]{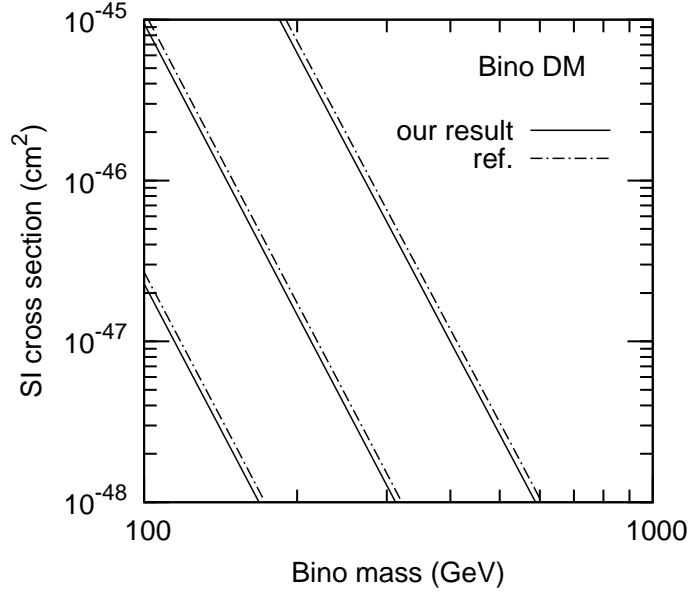} 
   \caption{$\tilde{\chi}$-$p$ SI scattering cross section as a
     function of neutralino mass.  Here we take parameters as
     $(m_{\tilde{q}L}/M$, $m_{\tilde{q}R}/M$, $\tilde{m}_{qLR}/M)$ =
     (1.5,~1.2,~0.1), and (2.0,~1.5,~1.0), and (3.0,~2.0,~1.0) from
     top to bottom. Solid lines show our result, while dashed-dot
     lines are the result in the case where light quark contribution
     in long-distance diagrams is included for reference. }
\label{fig:Binosigma}
 \end{center}
\end{figure}

Now we numerically show the cross section for the SI scattering
of the bino-like neutralino with nucleon. The interaction
of the bino-like neutralino with quark and squark is given as 
\begin{eqnarray}
{\cal L} &=&
\sqrt{2}g_Y
\sum_{q,i} \bar{q}
\left(c_{qL}^{(i)}P_R+c_{qR}^{(i)}P_L\right)
\tilde{\chi} \tilde{q}_i+{\rm h.c.} \ ,
\end{eqnarray}
where $P_{L/R}=(1\mp\gamma_5)/2$ and $g_Y$ is the gauge coupling constant
of $U(1)_Y$.  $\tilde{q}_1$ and $\tilde{q}_2$ are the lighter and
heavier squark, respectively, and the coefficients, $c_{qL}^{(i)}$ and
$c_{qR}^{(i)}$ $(i=1,2)$, are
\begin{eqnarray}
c_{qL}^{(1)}=-Y_{q_L} \cos\theta_{\tilde{q}} \ , &&
c_{qR}^{(1)}= Y_{q_R} \sin\theta_{\tilde{q}} \ , \nonumber\\
c_{qL}^{(2)}= Y_{q_L} \sin\theta_{\tilde{q}} \ , &&
c_{qR}^{(2)}= Y_{q_R} \cos\theta_{\tilde{q}} \ , 
\end{eqnarray}
where $Y_{u_L}=Y_{d_L}=1/6$, $Y_{u_R}= 2/3$ and $Y_{d_R}= -1/3$. The 
mixing angle of squark mass matrix $\theta_{\tilde{q}}$ is 
\begin{eqnarray}
  \tan2\theta_{\tilde{q}}&=&
  -\frac{2 m_q \tilde{m}_{qLR}}{m^2_{\tilde{q}L}-m^2_{\tilde{q}R}}\ , 
\end{eqnarray}
where we parametrize mass terms of left-handed and right-handed squark
($\tilde{q}_L$ and $\tilde{q}_R$) as
\begin{eqnarray}
{\cal L}^{\rm (mass)}_{\tilde{q}} =
-m_{\tilde{q}L}^2\tilde{q}_L^\ast \tilde{q}_L
-m_{\tilde{q}R}^2\tilde{q}_R^\ast \tilde{q}_R
+(m_q\tilde{m}_{qLR}\tilde{q}_L^\ast \tilde{q}_R + {\rm h.c.}) \ .
\end{eqnarray}
From the interaction Lagrangian, $a_q$ and $b_q$ can be written as
\begin{eqnarray}
  a_q^{(1)} = \frac{g_Y}{\sqrt{2}}
  (-Y_{qL}\cos \theta_{\tilde{q}}+Y_{qR}\sin \theta_{\tilde{q}})\ , &&
  b_q^{(1)} = \frac{g_Y}{\sqrt{2}}
  (-Y_{qL}\cos \theta_{\tilde{q}}-Y_{qR}\sin \theta_{\tilde{q}})\ , 
  \nonumber \\
  a_q^{(2)} = \frac{g_Y}{\sqrt{2}}
  (Y_{qL}\sin \theta_{\tilde{q}}+Y_{qR}\cos \theta_{\tilde{q}})\ ,  &&
  b_q^{(2)} = \frac{g_Y}{\sqrt{2}}
  (Y_{qL}\sin \theta_{\tilde{q}}-Y_{qR}\cos \theta_{\tilde{q}})\ .
\end{eqnarray}
Then, we numerically calculate the SI cross section, using
Eqs.~(\ref{eq:fG2}) and (\ref{treesq}).  (Here we include the QCD
correction to the long-distance contribution to $f_G$.)

\begin{figure}[t]
 \begin{center}
   \includegraphics[width=0.6\linewidth]{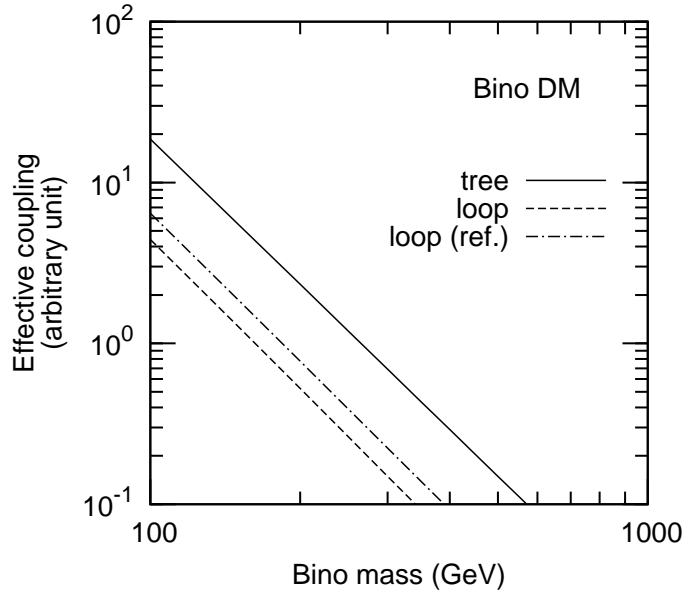} 
   \caption{Tree- and loop-level contributions in the effective
     coupling, $f_p$.  Here we take
     $(m_{\tilde{q}L}/m_{\tilde{\chi}}$,
     $m_{\tilde{q}R}/m_{\tilde{\chi}}$,
     $\tilde{m}_{qLR}/m_{\tilde{\chi}})$ = (3.0,~2.0,~1.0).  In the
     figure, tree- and loop-level ones are given in solid and dashed
     lines, respectively.  Here, we also give a dashed-dot line for
     the case where contribution of light quarks in long-distance
     loops is added in $f_G$ for reference.}
\label{fig:BinoDMfp}
 \end{center}
\end{figure}

The cross section for SI scattering with proton,
$\sigma_{\rm SI}^p$, is plotted in Fig.~\ref{fig:Binosigma} as a
function of the neutralino mass.  Here we give several results
in the figure by taking $(m_{\tilde{q}L}/M$, $m_{\tilde{q}R}/M$,
$\tilde{m}_{qLR}/M)$ = (1.5,~1.2,~0.1), (2.0,~1.5,~1.0), and
(3.0,~2.0,~1.0) from top to bottom in solid lines.  It is found that
the cross section is suppressed as the neutralino mass
become larger, and so it is unless masses of the neutralino
and squarks are degenerate.
In Fig.~\ref{fig:BinoDMfp}, we present tree- and loop-level
contributions to SI coupling in solid and dashed lines, respectively.
The tree-level contribution, which mainly comes from $g_q^{(1)}$ in
Eq.~(\ref{treesq}), is dominant, and the quark-squark loop
contribution is subdominant; it accounts for $O(10)$~\% in the SI cross
section.  Just for reference, we also give the contribution from loop
diagram in the case where the light quark contribution of
long-distance is added in $f_G$ (dashed-dot lines in
Figs.~\ref{fig:Binosigma} and \ref{fig:BinoDMfp}).  We have seen that
the difference between our evaluation and the reference one is
$O(10)$~\%.   As shown in Fig.~\ref{fig:Binosigma}, the result is
  almost independent of the choice of the squark masses if they are in
  the same order of bino mass.  This is because
    both $f_G^{\rm SD}|_q$ and $f_G^{\rm LD}|_q$ 
    give the same order of contribution to $f_G$ in such a
    case.\footnote{
     We note that there exist benchmark points to explain the
      observed dark matter abundance in which squark masses are much
      larger than the bino one.  In such a case, the lightest
      neutralino should be a mixed state of bino and Higgsinos so that
      the SI cross section is determined by the Higgs boson exchange.
      We have checked that the long-distance contribution induced by
      the Higgs exchange is correctly evaluated in the previous works
      ({\it e.g.} Ref.~\cite{Drees:1993bu}) and the numerical result
      of the cross section in such a case is unchanged. }
   The result, which has improved theoretical calculation of the SI
    cross section of dark matter with nucleon, will be important when
    we determine the local density and/or cross section of dark matter in the
    future experiments.
   Finally we also note that, in the public codes for studies of
  the neutralino dark matter, DarkSUSY \cite{darksusy} and MicroOMEGA
  \cite{Belanger:2008sj}, the gluon contribution to the spin-independent
  cross section is included by following
  Ref.~\cite{Drees:1993bu}. Those programs should be corrected when
  one needs to calculate the cross section at less than $O(10)$ \%
  error.

\section{Wino-like neutralino dark matter}
\label{sec:Wino}

In the anomaly mediation \cite{Randall:1998uk} the wino-like
neutralino is a candidate for dark matter in the universe. The thermal
relic abundance of the wino-like neutralino in the Universe is
consistent with the WMAP observation when the wino mass is from
2.7~TeV to 3.0~TeV \cite{Hisano:2006nn}.  The lighter wino-like
neutralino may be consistent with the observation of the dark matter
abundance in the Universe if decay of gravitino or other quasistable
particles may produce the dark matter nonthermally
\cite{Gherghetta:1999sw,Moroi:1999zb}.

The tree-level contribution to the cross section for elastic
scattering of the wino-like neutralino with nucleon is evaluated in
Ref.~\cite{Murakami:2000me}. However, in the case that the
superparticles except for gauginos and the heavier Higgs bosons are
much heavier than the weak scale in the MSSM, the tree-level
interactions of the neutralino to quarks are quite suppressed. The
scattering process is dominated by the $W$ boson loop diagrams.
Recently, we reevaluated the cross section for SI scattering of the
wino-like neutralino with nucleon \cite{Hisano:2010fy}. In the work,
the two-loop contribution to the effective interaction of the
neutralino with gluon is included, in addition to one-loop ones to the
interaction with quarks. Then it is confirmed that both one- and
two-loop contributions act as the leading contribution in the SI
scattering process.  Although calculation of two-loop diagrams is much
involved, it can be dealt with systematically in the Fock-Schwinger
gauge.  In the following, we show the technical detail of the
calculation.

\begin{figure}[t]
 \begin{center}
   \includegraphics[width=0.55\linewidth]{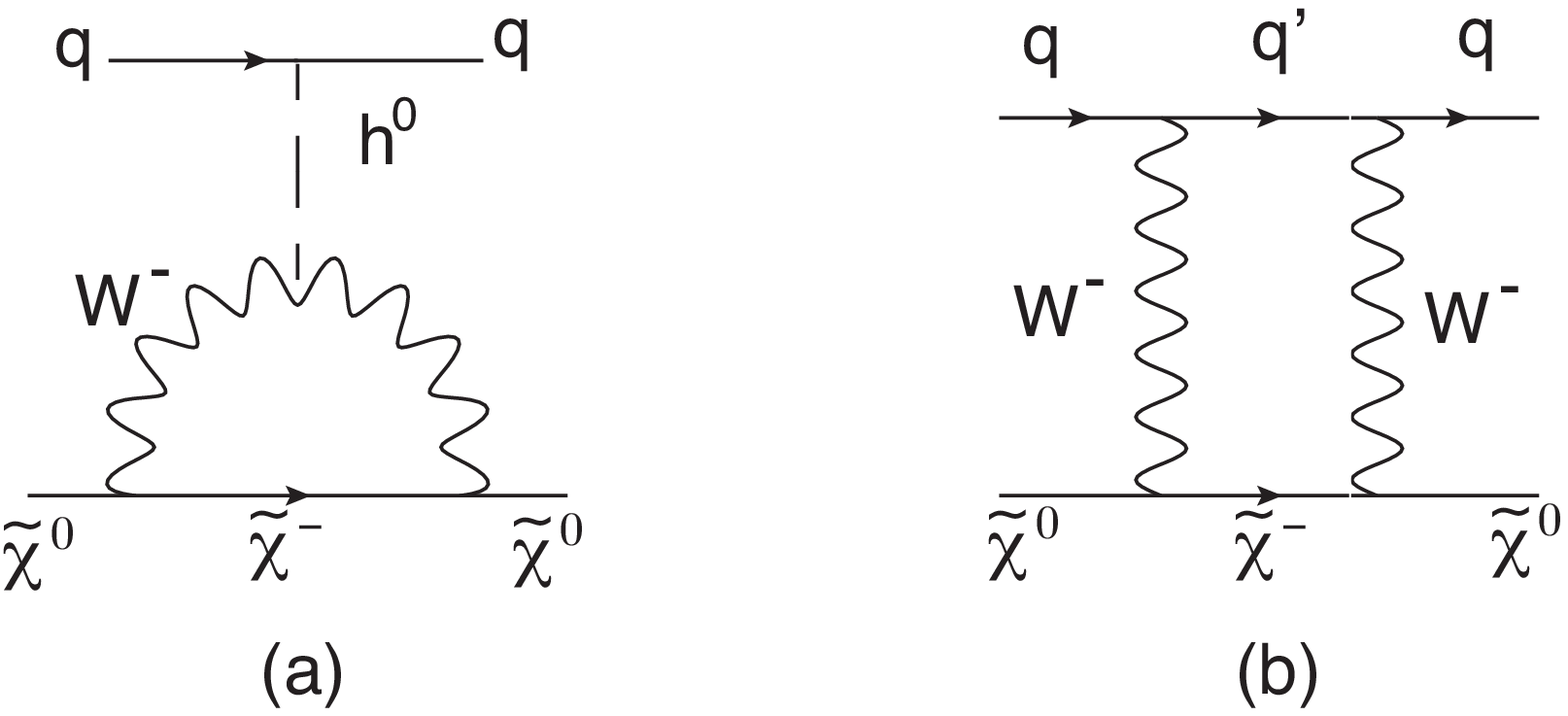} 
   \caption{One-loop contributions to effective interactions of
     the wino-like neutralino and light quarks.}
\label{fig:Wino1}
 \end{center}
\end{figure}
\begin{figure}[t]
 \begin{center}
   \includegraphics[width=0.7\linewidth]{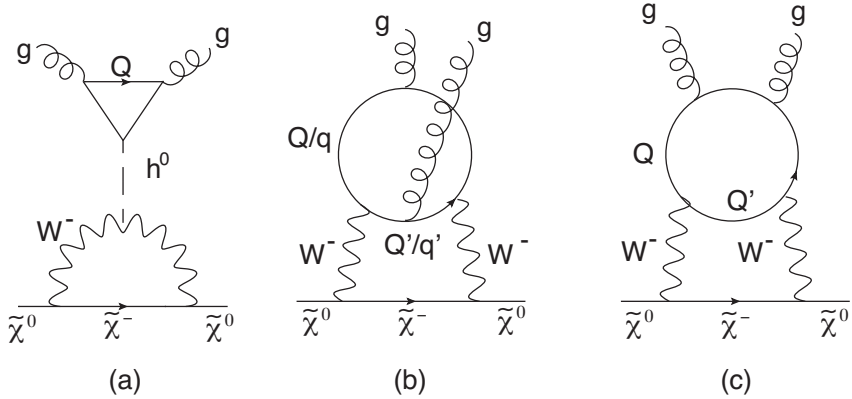} 
   \caption{Two-loop contributions to interactions of wino-like
     neutralino and gluon. }
\label{fig:Wino2}

 \end{center}
\end{figure}
Winos are gauginos which are partners of the $SU(2)_L$ gauge bosons in
the MSSM, and they have the weak interaction.  The relevant operators
in our discussion is 
\begin{eqnarray}
  {\cal L}
  &=&
  -g_2
  \left(
    \bar{\tilde{\chi}}\gamma^\mu\tilde{\chi}^-W^\dagger_\mu
    +
    {\rm h.c.}
  \right) \ ,
\end{eqnarray}
where $g_2$ is the gauge coupling constant of $SU(2)_L$.  The wino-like
neutralino accompanies the wino-like chargino ($\tilde{\chi}^-$).  The
mass difference is dominated by one-loop contribution unless the Higgsino
and wino masses are almost degenerate.  

Before getting down to the
two-loop calculation, we quickly review one-loop contribution to the
SI interaction of $\tilde{\chi}$-$N$. The effective interactions of
$\tilde{\chi}$ with light quarks in the effective
Lagrangian are generated by diagrams in Fig.~\ref{fig:Wino1}.  The
diagram (a), which is induced by the standard-model Higgs boson
($h^0$) exchange, contributes to $f_q$, while the diagram (b)
generates the other terms in the effective Lagrangian. The results are
given in Ref.~\cite{Hisano:2010fy}.  The obtained effective couplings
are finite in the limit $M\rightarrow \infty$.  Thus, the SI
interaction of $\tilde{\chi}$-$N$ is not suppressed even if the
wino-like neutralino is much larger than the $W$ boson mass, as
pointed out in Ref.~\cite{Hisano:2004pv}.

Next, let us derive the effective interaction of the neutralino with
gluon, which is generated at two-loop level. Three types of diagrams
in Fig.~\ref{fig:Wino2} contribute to $f_G$.  Diagram (a) includes
heavy quark loops, and it is evaluated from $f_Q$ as given
in Eq.~(\ref{matching2}).  On the other hand, we need to calculate
irreducible two-loop diagrams (b) and (c) explicitly.  In order to
handle this calculation, it is convenient to evaluate the vacuum
polarization tensor of $W$ boson in the gluon background,
\begin{eqnarray}
i \Pi_{\mu\nu} (q) &= & 
-\sum_{(u,d)}  \int \frac{d^4p}{(2\pi)^4}
\frac{g_2^2}{2}  ~{\rm Tr}\left[\gamma_\mu P_L S_d(p) 
P_R \gamma_\nu \tilde{S}_u(p-q)\right] \ .
\end{eqnarray}
Here, $\tilde{S}_u$ and $S_d$ are propagators of up-type and down-type
quarks, respectively, in the Fock-Schwinger gauge for gluon field.
The vacuum polarization tensor is decomposed to the transverse and
longitudinal parts as
\begin{eqnarray}
\Pi_{\mu\nu} (q) &\equiv & 
\Pi_T(q^2) (-g_{\mu\nu} +\frac{q_\mu q_\nu}{q^2})+
\Pi_L(q^2) \frac{q_\mu q_\nu}{q^2}
\ .
\end{eqnarray}
We found from an explicit calculation that the longitudinal part of
the self energy does not contribute to $f_G$. Thus, what we have to
calculate is coefficients for terms with gluon scalar operator
$G_{\mu\nu}^aG^{a\mu\nu}$ in $\Pi_T(q^2)$, {\it i.e.} $\Pi_T(q^2)|_{GG}$.

For the calculation $\Pi_T(q^2)|_{GG}$, we furthermore decompose the
contributions from diagram (b) and (c) in Fig.~\ref{fig:Wino2} as
follows;
\begin{eqnarray}
\Pi^{(b)}_T(q^2;i)|_{GG} &=& \frac{\alpha_2 \alpha_s}{24}
\left[
  6 B_{20}^{(2,2)}+q^2 B_{22}^{(2,2)}+q^2B_{1}^{(2,2)}
\right], \nonumber \\
\Pi^{(c1)}_T(q^2;i)|_{GG} &=& \frac{\alpha_2 \alpha_s}{4} m_u^2
\left[
  -2 B_{20}^{(4,1)}-q^2 B_{22}^{(4,1)}-q^2B_{1}^{(4,1)}
\right], \nonumber \\
\Pi^{(c2)}_T(q^2;i)|_{GG} &=& \frac{\alpha_2 \alpha_s}{4} m_d^2
\left[
  -2 B_{20}^{(1,4)}-q^2 B_{22}^{(1,4)}-q^2B_{1}^{(1,4)}
\right]\ ,
\end{eqnarray}
where $i$ is the flavor index and $\alpha_2=g^2_2/4 \pi$. $\Pi_T^{(b)}$ is
obtained from diagram (b) and $\Pi_T^{(c1)}$ ($\Pi_T^{(c2)}$) is
from diagram (c) in which gluon lines are attached with the
propagator of up- (down-) type quark.  In the above expressions, loop
integrals are defined as
\begin{eqnarray}
\int \frac{d^4q}{i \pi^2} \frac{q_{\mu}}{((p+q)^2-m_u^2)^n(q^2-m_{d}^2)^m} 
&\equiv& B_1^{(n,m)}p_{\mu}\ ,\nonumber\\
\int \frac{d^4q}{i \pi^2} \frac{q_\mu q_{\nu}}{((p+q)^2-m_u^2)^n
(q^2-m_{d}^2)^m} 
&\equiv& B_{20}^{(n,m)} g_{\mu\nu}+B_{22}^{(n,m)}p_{\mu}p_{\nu} \ ,
\label{integ}
\end{eqnarray}
where $m_u$ and $m_d$ are masses of up-type and down-type quarks,
respectively.

Notice that, in the loop integrals for the self-energy part of 
diagram (c), the typical momentum is the mass of quark which emit two
gluons, {\it i.e.} the up- and down-type quark mass in $\Pi_T^{(c1)}$ and
$\Pi_T^{(c2)}$, respectively.  This fact means that diagram (c)
gives the long-distance contribution in the part of one-loop self-energy
of the $W$ boson.  Therefore, we add the contribution of $\Pi_T^{(c1)}$
for $i=2$ and 3 (up-type quark is heavy one) and $\Pi_T^{(c2)}$ for
$i=3$ (down-type quark is heavy one).  On the other hand, the loop
momentum of quark loop in diagram (b) is dominated by the external
momentum of the quark loop diagram; therefore, all quarks contribute
in the loop. Then $\Pi_T(q^2)|_{GG}$ is properly obtained as
\begin{eqnarray}
\Pi_T(q^2)|_{GG} = \sum_{i=1,2,3} \Pi^{(b)}_T(q^2;i)|_{GG}
+ \sum_{i=2,3} \Pi^{(c1)}_T(q^2;i)|_{GG} + \Pi^{(c2)}_T(q^2;3)|_{GG} \ .
\end{eqnarray}
Executing the integral and neglecting quark masses except for the top
one, we finally get
\begin{eqnarray}
\Pi_T(q^2)|_{GG} 
&=&
\alpha_s \alpha_2
\left(\frac1{24} \frac{2q^2-3 m_t^2}{(q^2-m_t^2)^2}+2\times 
\frac1{12} \frac1{q^2}\right) \ .
\label{pit}
\end{eqnarray}
The first term in the right-handed side comes from loops of the third
generation, while the second term is induced by the first- and 
second-generation quarks.  We observed that the singular terms in $1/m^2_{q}$
are cancelled in the loop integrals of $\Pi_T^{(c1,c2)}$ so that
$\Pi_T^{(c1,c2)}$ vanishes in the $m_q\rightarrow 0$ limit.  Actually,
this is expected from the indication of Eq.~(\ref{eq:checkmatching})
and the fact that diagram (b) in Fig.~\ref{fig:Wino1} does not
contribute to the neutralino-quark scalar coupling $f_q$ in a limit of
zero quark mass \cite{Hisano:2010fy}.

Combining the above results, we get $f_G$ as
\begin{eqnarray}
f_G = -(c_c+c_b+c_t)\times \frac{\alpha_s}{12\pi} 
\frac{\alpha_2^2}{4m_W m_{h^0}^2} g_{\rm H}(x)
+\frac{\alpha_s}{4\pi} 
\frac{\alpha_2^2}{m_W^3} g_{\rm B3}(x,y)
+2\times \frac{\alpha_s}{4\pi} 
 \frac{\alpha_2^2}{m_W^3} g_{\rm B1}(x) \ ,
\end{eqnarray}
where $m_{h^0}$ and $m_W$ are the Higgs and $W$-boson masses,
respectively, and $y=m_t^2/M^2$.  In addition, we include the QCD
correction to the long-distance contribution of heavy quarks to $f_G$.
The mass functions $g_H(x)$, $g_{\rm B3}(x,y)$ and $g_{\rm B1}(x)$ are
given in Ref.~\cite{Hisano:2004pv}.

\begin{figure}[t]
 \begin{center}
   \includegraphics[width=0.6\linewidth]{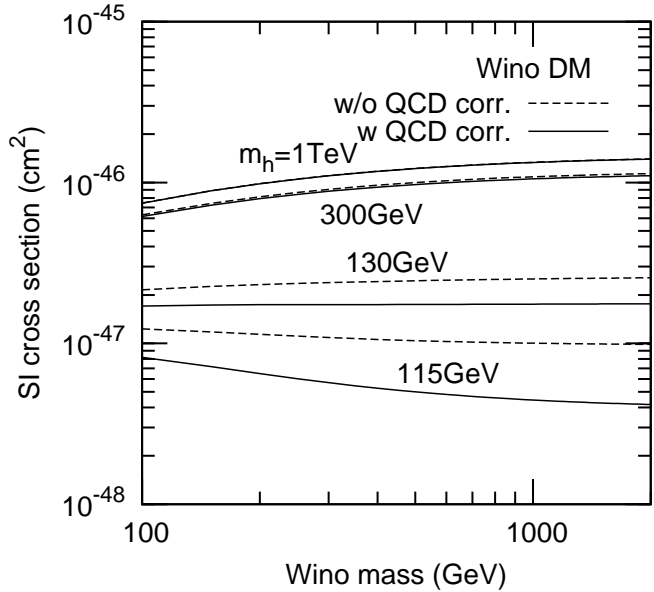} 
   \caption{$\tilde{\chi}$-$p$ SI scattering cross section as a
     function of neutralino mass $M$ for wino-like neutralino case.
     Here we take the Higgs boson mass as 115, 130, 300 GeV, and 1 TeV
     from top to bottom.  Results where the QCD correction is taken into
     account for long-distance contribution are given in solid lines,
     and those without the QCD correction are in dashed lines. 
\label{fig:Winosigma}
}
 \end{center}
\end{figure}

In Fig.~\ref{fig:Winosigma}, we show the numerical result for the SI cross
section.  Here the evaluation with (without) the QCD correction is
given by taking $m_{h^0}=115,~130$, $300~{\rm GeV}$, and $1~{\rm
  TeV}$. While the latter two values may not be realistic in
  the MSSM, the next-minimal supersymmetric standard model, for
  example, may predict a larger Higgs boson mass.  It is found from the
figure that the cross section is more suppressed unless the Higgs boson
mass is larger than a few hundred GeV.  This fact is the consequence
of an accidental cancellation in SI coupling as it is shown in
Ref.~\cite{Hisano:2010fy}; the QCD correction makes the contribution from
the Higgs boson exchange larger so that the cancellation works
hard.\footnote{
   In Ref.~\cite{Hisano:2010fy}, the QCD correction for
  long-distance contribution and the contribution from the twist-2
  operators of bottom and charm quarks are not included.
}
On the other hand, the effect of the cancellation becomes weaker
as the Higgs boson mass is larger.

\section{Conclusion}
\label{sec:conc}

In this article, we have systematically calculated the spin-independent
cross section of the nucleon-lightest neutralino scattering process at
loop level.  Although the one-loop (higher-loop) diagrams are usually
the higher-order of quantum correction to tree-level (one-loop) ones,
such higher order diagrams of neutralino-gluon scattering work as the
leading contribution.  We have evaluated the effective coupling of
neutralino-gluon scattering by explicit calculation of loop diagrams
in the Fock-Schwinger gauge.  In this gauge, we can calculate
systematically the loop diagrams, tracking the loop momentum scales
which dominate in the integrals.  Making the most of it, we have
separately evaluated short- and long-distance contributions in the
loop diagrams, {\it i.e.} the contributions in which the loop momentum
is characterized by masses of new particles heavier than quarks (like
mass scale of neutralino or squark) and those of quarks.  Being aware
that the contribution of light quarks must not be included in the
long-distance diagrams, we have numerically calculated the
spin-independent effective coupling to give the cross section of
nucleon-neutralino scattering.  In order to focus on the contribution
of neutralino-gluon scattering, we considered the bino- and wino-like
neutralino cases.  In the former case, we have found that the
effective coupling of neutralino-gluon scattering differs by 20-30~\%
from the one evaluated in the case where light quarks are included in
the long-distance contribution incorrectly; it gives $O(10)~\%$
alteration in the SI cross section.  This improvement in the
  theoretical calculation of the cross section will be essential when
  we explore the nature ({\it i.e.} interaction and/or local density) of dark
  matter in the future experiments.
On the other hand, in the latter case, the quark- and gluon-neutralino
scattering contributes in almost the same magnitude and opposite
sign. Then, they cancel each other to suppress the SI cross section.
Here we have also taken into account QCD correction for long-distance
contribution; consequently, we have found that the cancellation works
hard.



\section*{Acknowledgment}

The work was supported in part by the Grant-in-Aid for the Ministry of
Education, Culture, Sports, Science, and Technology, Government of
Japan, No. 20244037, No. 2054252 and No. 2244021 (J.H.) and Research
Fellowships of the Japan Society for the Promotion of Science for
Young Scientists (K.I.).  The work of J.H. is also supported by the
World Premier International Research Center Initiative (WPI
Initiative), MEXT, Japan.


\appendix

\section{Propagators in Fock-Schwinger gauge}
Here we present colored fermion and scalar boson propagators in the
Fock-Schwinger gauge for gluon, and useful formulas to evaluate effective 
coupling of WIMP and gluon. We also show derivation of the coupling 
in a model in which the WIMP has an interaction with the Higgs boson, as a
simple exercise. 

Defining gluon field $A^a_{\mu}$ in covariant derivative,
$D_{\mu}\equiv \partial_\mu + i g_s A^a_{\mu}T_a$, the gauge-fixing
condition in the Fock-Schwinger gauge is written as
\begin{eqnarray}
x^\mu A_\mu^a(x)&=&0 \ .
\end{eqnarray}
In this gauge, the vector field can be directly expressed in the terms of
the gluon field strength tensor as
\begin{eqnarray}
A_\mu^a(x)&=&\int^1_0 d\alpha ~\alpha x^\rho ~G^a_{\rho\mu}(\alpha x)
\nonumber\\
&=&
\frac{1}{2\cdot 0 !} x^\rho G^a_{\rho\mu}(0)
+
\frac{1}{3\cdot 1 !} 
x^\alpha x^\rho (D_\alpha  G_{\rho\mu}(0))^a
+
\frac{1}{4\cdot 2 !} 
x^\alpha x^\beta x^\rho (D_\alpha D_\beta G_{\rho\mu}(0))^a
+
\cdots.\nonumber\\
&&
\end{eqnarray}
While the gauge-fixing condition breaks the translation invariance, it
is recovered in the gauge-invariant objects. The propagators of colored fermion
in the gluon background are given as
%
\begin{figure}[t]
 \begin{center}
   \includegraphics[width=0.7\linewidth]{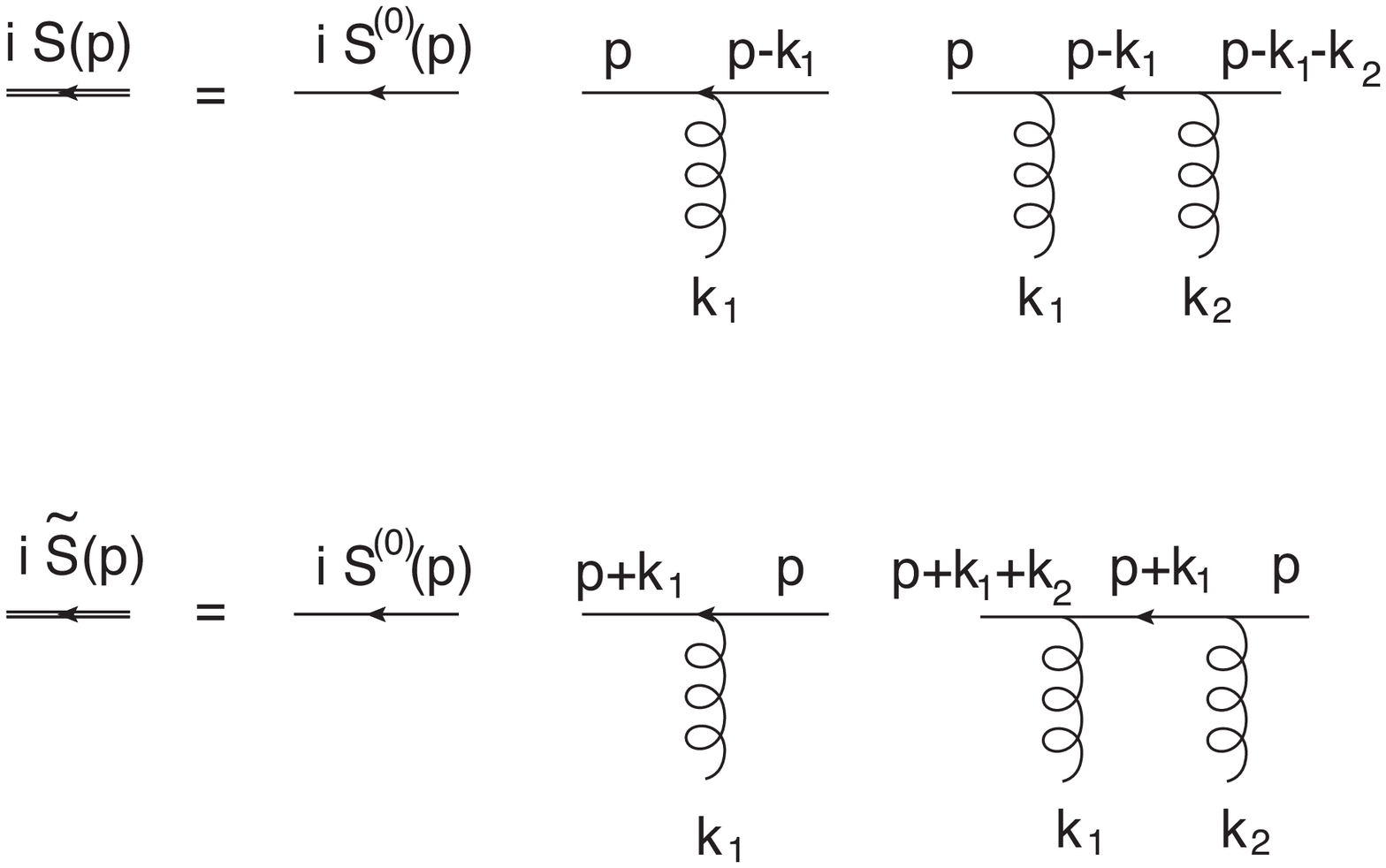} 
   \caption{Fermion propagators.}
 \end{center}
\end{figure}
\begin{eqnarray}
i S(p) &\equiv& \int 
d^4 x ~e^{i p x} 
\langle T\{\psi(x)\bar{\psi}(0)\}\rangle
\nonumber\\
&=& i S^{(0)}(p) \nonumber\\
&+& \int d^4 k_1 ~ i S^{(0)}(p)~g_s\gamma^\alpha
\left(\frac12 G_{\alpha\mu}  
\frac{\partial}{\partial k_{1\mu}} 
\delta^{(4)}(k_1)\right)
i S^{(0)}(p-k_1) \nonumber\\
&+& \int d^4 k_1d^4 k_2 ~ i S^{(0)}(p) ~g_s\gamma^\alpha
\left(\frac12 G_{\alpha\mu}  
\frac{\partial}{\partial k_{1\mu}} 
\delta^{(4)}(k_1)\right)
i S^{(0)}(p-k_1) ~g_s\gamma^\beta
\nonumber\\
&&\times 
\left(\frac12 G_{\beta\nu}  
\frac{\partial}{\partial k_{2\nu}} 
\delta^{(4)}(k_2)\right)
i S^{(0)}(p-k_1-k_2) + \cdots ~,
\\
i \tilde{S}(p) &\equiv& \int 
d^4 x~ e^{-i p x} 
\langle T\{\psi(0)\bar{\psi}(x)\}\rangle
\nonumber\\
&=& i S^{(0)}(p) \nonumber\\
&+& \int d^4 k_1 ~ i S^{(0)}(p+k_1)~g_s\gamma^\alpha
\left(\frac12 G_{\alpha\mu}  
\frac{\partial}{\partial k_{1\mu}} 
\delta^{(4)}(p)\right)
i S^{(0)}(p) \nonumber\\
&+& \int d^4 k_1d^4 k_2 ~ i S^{(0)}(p+k_1+k_2)~g_s\gamma^\alpha
\left(\frac12 G_{\alpha\mu}  
\frac{\partial}{\partial k_{2\mu}} 
\delta^{(4)}(k_2)\right)
\nonumber\\
&&\times 
i S^{(0)}(p+k_1)~g_s \gamma^\beta
\left(\frac12 g G_{\beta\nu}  
\frac{\partial}{\partial k_{1\nu}} 
\delta^{(4)}(k_1)\right)
i S^{(0)}(p) + \cdots ~,
\end{eqnarray}
where $i S^{(0)}(p)=i/(\sla{p}-m)$ and $G_{\mu\nu}\equiv G_{\mu\nu}^aT_a$.  In the actual calculation, terms
including covariant derivatives, such as $D_\alpha G_{\mu\nu}$ and
$D_\alpha D_\beta G_{\mu\nu}$, are ignored, since they are irrelevant
to evaluation of the SI cross section. In text we evaluate the scalar
coupling of WIMP with gluon, $f_G$, by extracting out the bilinear
term of the gluon field strength from the WIMP ($W$ boson) two-point
function in the bino- (wino-)like neutralino dark matter case. The scalar
operator of gluon $G_{\mu\nu}^aG^{a\mu\nu}$ is projected out from the
bilinear term of the gluon field strength as
\begin{eqnarray}
G_{\alpha\mu}^aG_{\beta\nu}^a
&=&
\frac{1}{12}
G_{\rho\sigma}^aG^{a\rho\sigma}
(g_{\alpha\beta}g_{\mu\nu}
-g_{\alpha\nu}g_{\beta\mu}) 
\nonumber\\
&-&\frac12 g_{\alpha\beta} {\cal O}_{\mu\nu}^g
-\frac12 g_{\mu\nu} {\cal O}_{\alpha\beta}^g
-\frac12 g_{\alpha\nu} {\cal O}_{\beta\mu}^g
-\frac12 g_{\beta\mu} {\cal O}_{\alpha\nu}^g
\nonumber\\
&+& {\cal O}_{\alpha\mu\beta\nu}^g \ ,
\end{eqnarray}
where ${\cal O}_{\mu\nu}^g$ is the twist-2 operator of gluon
(Eq.~(\ref{twist2})) and ${\cal O}_{\alpha\mu\beta\nu}^g$ is a
higher-twist operator,
 \begin{eqnarray}
 {\cal O}_{\alpha\mu\beta\nu}^g &\equiv&
   G_{\alpha\mu}^aG_{\beta\nu}^a
   \nonumber\\
   &-&\frac12 g_{\alpha\beta} G_{~~\mu}^{a\rho}G_{\rho\nu}^{a}
   -\frac12 g_{\mu\nu} G_{~~\alpha}^{a\rho}G_{\rho\beta}^{a} +\frac12
   g_{\alpha\nu} G_{~~\beta}^{a\rho}G_{\rho\mu}^{a} +\frac12
   g_{\beta\mu} G_{~~\alpha}^{a\rho}G_{\rho\nu}^{a}
   \nonumber\\
   &+&\frac16 G_{\rho\sigma}^aG^{a\rho\sigma}
   (g_{\alpha\beta}g_{\mu\nu} -g_{\alpha\nu}g_{\beta\mu}) \ .
\end{eqnarray}

We also give colored scalar propagators:
\begin{eqnarray}
i \Delta(p) &\equiv& \int 
d^4 x ~e^{i p x} 
\langle T\{\phi(x)\phi^\dagger(0)\}\rangle
\nonumber\\
&=& i \Delta^{(0)}(p) \nonumber\\
&+& \int d^4 k_1 ~ i \Delta^{(0)}(p)
~g_s(2p-k_1)^\alpha
\left(\frac12  G_{\alpha\mu}  
\frac{\partial}{\partial k_{1\mu}} 
\delta^{(4)}(k_1)\right)
i \Delta^{(0)}(p-k_1) \nonumber\\
&+& \int d^4 k_1d^4 k_2 ~ i \Delta^{(0)}(p)
~g_s(2p-k_1)^\alpha
\left(\frac12  G_{\alpha\mu}  
\frac{\partial}{\partial k_{1\mu}} 
\delta^{(4)}(k_1)\right)
i \Delta^{(0)}(p-k_1) 
\nonumber\\
&&\times 
g_s(2p-2k_1-k_2)^\beta
\left(\frac12 G_{\beta\nu}  
\frac{\partial}{\partial k_{2\nu}} 
\delta^{(4)}(k_2)\right)
i \Delta^{(0)}(p-k_1-k_2)
\nonumber \\
&+& \int d^4 k_1d^4 k_2 ~ i \Delta^{(0)}(p)
(-ig_s^2)
\left(\frac12 G_{\alpha\mu}  
\frac{\partial}{\partial k_{1\mu}} 
\delta^{(4)}(k_1)\right)
\left(\frac12 G^\alpha_{~\nu}  
\frac{\partial}{\partial k_{2\nu}} 
\delta^{(4)}(k_2)\right)
\nonumber \\
&&\times 
i \Delta^{(0)}(p-k_1-k_2)\ ,
\end{eqnarray}
\begin{eqnarray}
i \tilde{\Delta}(p) &\equiv& \int 
d^4 x ~e^{-i p x} 
\langle T\{\phi(0)\phi^\dagger(x)\}\rangle
\nonumber\\
&=& i \Delta^{(0)}(p) \nonumber\\
&+& \int d^4 k_1 ~ i \Delta^{(0)}(p+k_1)
~g_s(2p+k_1)^\alpha
\left(\frac12  G_{\alpha\mu}  
\frac{\partial}{\partial k_{1\mu}} 
\delta^{(4)}(k_1)\right)
i \Delta^{(0)}(p) \nonumber\\
&+& \int d^4 k_1d^4 k_2 ~ i \Delta^{(0)}(p+k_1+k_2)
~g_s(2p+2k_1+k_2)^\alpha
\left(\frac12 G_{\alpha\mu}  
\frac{\partial}{\partial k_{1\mu}} 
\delta^{(4)}(k_2)\right)
\nonumber\\
&&\times
i \Delta^{(0)}(p+k_1) ~
g_s (2p+k_1)^\beta 
\left(\frac12 G_{\beta\nu}  
\frac{\partial}{\partial k_{2\nu}} 
\delta^{(4)}(k_1)\right)
i \Delta^{(0)}(p)
\nonumber \\
&+& \int d^4 k_1d^4 k_2 ~ i \Delta^{(0)}(p+k_1+k_2)
\nonumber \\
&&\times 
(-ig_s^2)
\left(\frac12  G_{\alpha\mu}  
\frac{\partial}{\partial k_{1\mu}} 
\delta^{(4)}(k_1)\right)
\left(\frac12  G^\alpha_{~\nu}  
\frac{\partial}{\partial k_{2\nu}} 
\delta^{(4)}(k_2)\right)
i \Delta^{(0)}(p)\ ,
\end{eqnarray}
%
\begin{figure}[t]
 \begin{center}
   \includegraphics[width=0.9\linewidth]{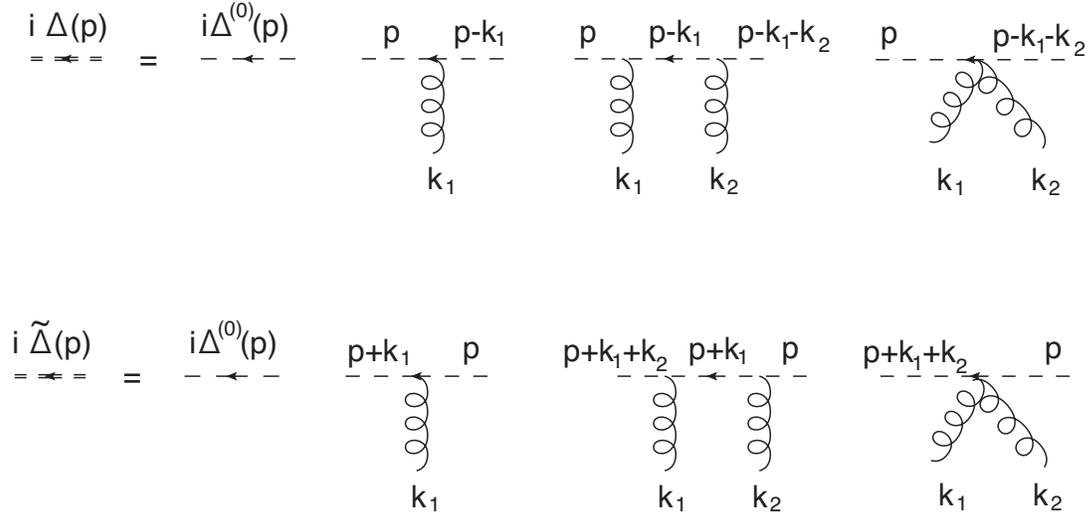} 
   \caption{Scalar boson propagators.}
 \end{center}
\end{figure}
%
where $i \Delta^{(0)}(p)=i/(p^2-m^2)$. The scalar propagators are 
reduced in a more convenient form for practical usage as
\begin{eqnarray}
i \Delta(p) 
&=& i \Delta^{(0)}(p) \nonumber\\
&+& \int d^4 k_1 ~ 
\left(\frac{\partial}{\partial k_{1\mu}} 
\delta^{(4)}(k_1) \right)
g p^{\alpha} G_{\alpha\mu} (i \Delta^{(0)}(p))^2
\nonumber \\
&+& \int d^4 k_1d^4 k_2 ~  
\left(\frac{\partial}{\partial k_{1\mu}} 
\delta^{(4)}(k_1) \right)
\left(\frac{\partial}{\partial k_{2\nu}} 
\delta^{(4)}(k_2) \right)
g^2 p^{\alpha}p^{\beta} G_{\alpha\mu} G_{\beta\nu}
(i \Delta^{(0)}(p))^3
\nonumber \\
&-& \int d^4 k_2
\left(\frac{\partial}{\partial k_{2\nu}} 
\delta^{(4)}(k_2) \right)
g^2 p^{\alpha}G_{\alpha\mu}G^{\mu}_{~\nu}
(i \Delta^{(0)}(p))^3
\nonumber \\
&+& \int d^4 k_1d^4 k_2 ~ i \Delta^{(0)}(p)
(-ig_s^2)
\left(\frac12 G_{\alpha\mu}  
\frac{\partial}{\partial k_{1\mu}} 
\delta^{(4)}(k_1)\right)
\left(\frac12 G^\alpha_{~\nu}  
\frac{\partial}{\partial k_{2\nu}} 
\delta^{(4)}(k_2)\right)
\nonumber \\
&&\times 
i \Delta^{(0)}(p-k_1-k_2) \ ,
\\
i \tilde{\Delta}(p) 
&=& i \Delta^{(0)}(p) \nonumber\\
&+& \int d^4 k_1 ~ 
\left(\frac{\partial}{\partial k_{1\mu}} 
\delta^{(4)}(k_1) \right)
g  p^{\alpha} G_{\alpha\mu} (i \Delta^{(0)}(p))^2
\nonumber \\
&+& \int d^4 k_1d^4 k_2 ~  
\left(\frac{\partial}{\partial k_{1\mu}} 
\delta^{(4)}(k_1) \right)
\left(\frac{\partial}{\partial k_{2\nu}} 
\delta^{(4)}(k_2) \right)
g^2 p^{\alpha}p^{\beta} G_{\alpha\mu} G_{\beta\nu}
(i \Delta^{(0)}(p))^3
\nonumber \\
&+& \int d^4 k_2
\left(\frac{\partial}{\partial k_{2\nu}} 
\delta^{(4)}(k_2) \right)
g^2 p^{\alpha}G_{\alpha\mu}G^{\mu}_{~\nu}
(i \Delta^{(0)}(p))^3
\nonumber \\
&+& \int d^4 k_1d^4 k_2 ~ i \Delta^{(0)}(p+k_1+k_2)
\nonumber \\
&&\times 
(-ig_s^2)
\left(\frac12 G_{\alpha\mu}  
\frac{\partial}{\partial k_{1\mu}} 
\delta^{(4)}(k_1)\right)
\left(\frac12 G^\alpha_{~\nu}  
\frac{\partial}{\partial k_{2\nu}} 
\delta^{(4)}(k_2)\right)
i \Delta^{(0)}(p)\ .
\end{eqnarray}
In the calculation of the SI cross section of the bino-like neutralino
in Sec.~\ref{sec:Bino}, the diagrams (a) and (c) in Fig.~\ref{fermion_dm}
are found to be zero in the Fock-Schwinger gauge, since terms except the first 
and fifth terms in $i \Delta(p) $ and $i \tilde{\Delta}(p)$ vanish in the calculation. 

\begin{figure}[t]
 \begin{center}
   \includegraphics[width=0.5\linewidth]{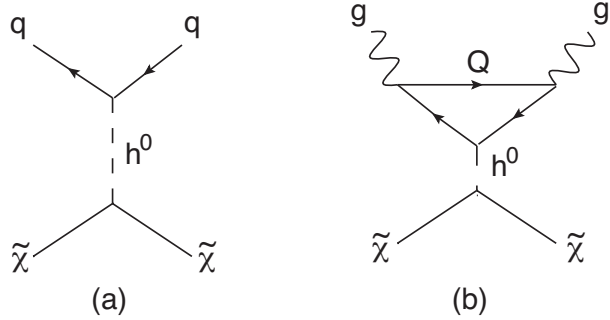} 
   \caption{Effective interactions of $\tilde{\chi}$ with light quarks and 
gluon, which are induced by Higgs boson exchange. }
\label{fig:higgs}
 \end{center}
\end{figure}

Last, we show the loop diagram calculation in the Fock-Schwinger gauge for
a simple scenario as an exercise.  We consider a model in which the
WIMP $\tilde{\chi}$ interacts with the standard-model Higgs boson;
\begin{eqnarray}
{\cal L} &=&-  c ~\bar{\tilde{\chi}}\tilde{\chi} h^0 \ .
\end{eqnarray}
Effective interactions of $\tilde{\chi}$ with light (and heavy) quarks
generated by diagram (a) in Fig.~\ref{fig:higgs};
\begin{eqnarray}
f_{q(Q)} &=&\frac{c g_2}{2 m_W m_{h^0}^2} \ ,
\end{eqnarray}
at leading order.  Other coefficients in
Eqs.~(\ref{eff_lagq}) and (\ref{eff_lagg}) are zero at the leading order.
The effective coupling of WIMP with gluon is induced by
diagram (b) in Fig.~\ref{fig:higgs};
\begin{eqnarray}
f_G&=& -\sum_{Q=c,b,t} f_Q m_Q
\int \frac{d^4p}{(2\pi)^4} {\rm Tr}[iS(p)]\left|_{GG} \right. \nonumber \\
&=& \frac{\alpha_s}{4\pi} \sum_{Q=c,b,t}
 f_Q m_Q \int \frac{d^4p}{i\pi^2}\frac{m_Q p^2}{(p^2-m_Q^2)^4} 
\nonumber \\
&=&
-  \frac{\alpha_s}{12 \pi} \times 3f_Q.
\label{fghiggs}
\end{eqnarray}
Here we neglected the QCD correction just for simplicity.

{}

\end{document}